\documentclass[a4paper]{article}
\usepackage{float}
\usepackage{INTERSPEECH2018}
\usepackage{here,url,cite}

\DeclareMathOperator*{\argmax}{argmax}

\title{Building state-of-the-art distant speech recognition using the CHiME-4 challenge with a setup of speech enhancement baseline}
\name{Szu-Jui Chen, Aswin Shanmugam Subramanian, Hainan Xu, Shinji Watanabe}
\address{
  Center for Language and Speech Processing, Johns Hopkins University, Baltimore, MD 21218, USA}
\email{\{schen146,asubra13,hxu31,shinjiw\}@jhu.edu}

\begin{document}

\maketitle 
\begin{abstract}
This paper describes a new baseline system for automatic speech recognition (ASR) in the CHiME-4 challenge to promote the development of noisy ASR in speech processing communities by providing 1) state-of-the-art system with a simplified single system comparable to the complicated top systems in the challenge, 2) publicly available and reproducible recipe through the main repository in the Kaldi speech recognition toolkit.
The proposed system adopts generalized eigenvalue beamforming with bidirectional long short-term memory (LSTM) mask estimation.
We also propose to use a time delay neural network (TDNN) based on the lattice-free version of the maximum mutual information (LF-MMI) trained with augmented all six microphones plus the enhanced data after beamforming.
Finally, we use a LSTM language model for lattice and n-best re-scoring. 
The final system achieved 2.74\% WER for the real test set in the 6-channel track, which corresponds to the 2nd place in the challenge.
In addition, the proposed baseline recipe includes four different speech enhancement measures, short-time objective intelligibility measure (STOI), extended STOI (eSTOI), perceptual evaluation of speech quality (PESQ) and speech distortion ratio (SDR) for the simulation test set. 
Thus, the recipe also provides an experimental platform for speech enhancement studies with these performance measures.

\end{abstract}

\section{Introduction}
\label{sec:intro}
In recent years, multi-channel speech recognition has been applied on devices used in daily life, such as Amazon Echo and Google Home. 
The recognition accuracy is greatly improved by exploiting microphone arrays when compared to single channel microphone devices \cite{barker2015third,kinoshita2016summary,li2017acoustic}. 
However, satisfactory performance is still not achieved in noisy everyday environments. 
Hence, the CHiME-4 challenge is designed to conquer this scenario by recognizing speech in challenging noisy environments \cite{vincent2017analysis}.
Through the series of the challenge activities, several speech enhancement and recognition techniques are established as an effective method for this scenario including mask-based beamforming, multichannel data augmentation, and system combination with various front-end techniques \cite{yoshioka2015ntt,du2016ustc,menne2016rwth,erdogan2016multi,fujita2016unsupervised}.


Although many submitted systems in the CHiME-4 challenge have yielded a lot of outcomes in this multi-channel Automatic Speech Recognition (ASR) scenario \cite{du2016ustc,menne2016rwth,erdogan2016multi}, one of the drawbacks is that all top systems are highly complicated due to multiple systems and fusion techniques, and it is not easy for the other research groups to follow these outcomes.
This paper aims to deal with the above drawback by building a new baseline to promote the development of noisy ASR in speech enhancement, separation, and recognition communities.

\begin{figure}[t]
  \centering
  \includegraphics[width=\linewidth]{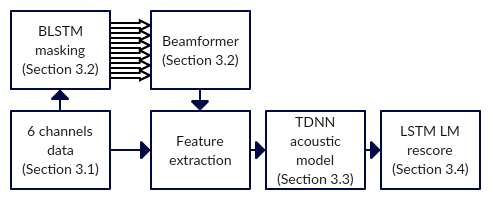}
  \caption{Diagram of speech recognition system.}
  \label{fig:speech_sys}
\end{figure}

We propose a single ASR system to further push the border of this challenge. 
Most important of all, our system is reproducible since it is implemented in the Kaldi ASR toolkit and other opensource toolkits.
All the scripts in our experiments can be downloaded from the official GitHub website\footnote{https://github.com/kaldi-asr/kaldi/pull/2142}.
The original CHiME-4 baseline is described in \cite{vincent2017analysis}, which uses a delay-and-sum beamformer (BeamformIt) \cite{anguera2007acoustic}, a deep neural network with state-level minimum Bayes Risk (DNN+sMBR) criterion \cite{vesely2013sequence}, and recurrent neural network-based language model (RNNLM) \cite{mikolov2010recurrent}. 
On the contrary, our proposed system is shown in Figure \ref{fig:speech_sys}. 
We adopt to use Bidirectional long short-term memory (BLSTM) mask based beamformer (Section \ref{sec:blstm}), which has been shown to be more effective \cite{erdogan2016improved,nn-gev} than BeamformIt. 
For an acoustic model, the DNN used in baseline is limited to represent long-term dependencies between acoustic characteristics. 
Hence, a sub-sampled time delay neural network (TDNN) \cite{waibel1990phoneme} with the lattice-free version of the maximum mutual information (LF-MMI) is used for our acoustic model \cite{povey2016purely} (Section \ref{sec:tdnn}).
This paper also shows the great improvement on the word error rate (WER) when we combine it with data augmentation in a multichannel scenario using all six microphones plus the enhanced data after beamforming.
Then, we further use a LSTM language model (LSTMLM), which uses a new training criterion and importance sampling, and has been shown to be more efficient and better in performance \cite{xuneural}, to re-score hypotheses.

We also incorporate computation of four different speech enhancement measures in our recipe - perceptual evaluation of speech quality (PESQ) \cite{pesq},  short-time objective intelligibility measure (STOI) \cite{stoi}, extended STOI (eSTOI) \cite{estoi} and speech distortion ratio (SDR) \cite{sdr}. 
We include these measurements as part of the recipe for two reasons. 
First, the ASR performance shows only one aspect of the speech enhancement algorithm. 
Objective enhancement metrics can give an indication on how well the enhancement is with different aspects (e.g., intelligibility, signal distortions).
Second, testing an enhancement algorithm with ASR takes a significant amount of computational time, whereas obtaining these scores is quite fast. 
Hence, it can give an initial indication of how good the enhancement is.

\section{Related work}
In \cite{du2016ustc}, a fusion system in the DNN posterior domain is proposed to get the best result in the competition. 
\cite{menne2016rwth,erdogan2016multi,fujita2016unsupervised} also use fusion systems in the decoding hypothesis domain with multiple systems mainly using different front-end techniques.
Unlike these highly complicated systems, our proposed system is based on a {\it single system} without the above fusion systems, yet achieves comparable performance to these top systems in the challenge task.
One of the unique technical aspects of our proposed system is to fully utilize the effectiveness of TDNN with LF-MMI by combining it with multichannel data augmentation techniques, which achieves significant improvement.
Our new LSTMLM also contributes to boost the final performance.
\label{sec:advance}

\section{Proposed system}
\noindent Our system starts from BLSTM mask based beamformer and followed by feature extraction. Phoneme to audio alignments are then generated by GMM acoustic model and are fed into TDNN acoustic model for training. Finally, the lattices after first pass decoding in TDNN is re-scored by a 5-gram LM and further re-scored by LSTMLM.

\subsection{Data augmentation}
\label{sec:6data}
Training with multichannel data has been shown to be effective for ASR systems \cite{erdogan2016multi,barker2015third,hori2017multi}. 
This augmentation can increase the variety in the training data and help the generalization to test set. 
In our work, we not only use data from all 6 channels but also add the enhanced data generated by beamformer to training set.

Let $O = (\mathbf{o} (t) \in \mathbb{R} ^D | t=1, \dots, T)$ be a sequence of $D$-dimensional feature vectors with length $T$, which is a single channel speech recognition case. 
In our case, we deal with an $M$-channel input ($M=6$), which is represented as $\mathbf{O} = (\mathbf{o} _{m} (t) \in \mathbb{R} ^D | t=1, \dots, T, m =1, \dots, M)$.
Then, the original training method only uses a particular channel input (e.g., $m$-th input) as training data to obtain acoustic model parameters $\Theta$, as follows:
\begin{equation}
 \hat{\Theta} = \arg \max _{\Theta} \mathcal{L} (\mathbf{O}_m),
\end{equation}
where $\mathcal{L}$ is an objective function (log likelihood for the GMM case and negative cross entropy for the DNN case), with reference labels as supervisions.
Data augmentation approach tries to use training data of all channels, as follows:
\begin{equation}
 \hat{\Theta} = \arg \max _{\Theta} \mathcal{L} (\mathbf{O} = \{\mathbf{O}_m\} _{m=1}^M)
\end{equation}

Further, we extend to include an enhanced data $\mathbf{O} ^{\text{enh}} = (\mathbf{o} ^{\text{enh} (t)} \in \mathbb{R} ^D | t=1, \dots, T)$ with the above multichannel data, that is
\begin{equation}
 \hat{\Theta} = \arg \max _{\Theta} \mathcal{L} (\{\mathbf{O}, \mathbf{O} ^{\text{enh}}\}),
\end{equation}
where the enhancement data $\mathbf{O} ^{\text{enh}}$ is obtained by a single-channel masking or beamformer method, which is described in Section \ref{sec:blstm}.

\subsection{BLSTM mask based beamformer}
\label{sec:blstm}
We use the BLSTM mask based Generalized Eigenvalue (GEV) beamformer described in \cite{nn-gev}. The GEV beamforming procedure requires an estimate of the Cross-Power Spectral Density (PSD) matrix of the noise and the target speech. The BLSTM model estimates two masks: the first mask indicates the time frequency bin that are probably dominated by speech and the other indicates which are dominated by noise. 
With the combined speech and noise masks, we can estimate the PSD matrices of speech components $\mathbf{\Phi}_{\text{speech}}(b) \in \mathbb{C} ^{M\times M}$ at frequency bin $b$, and that of noise components $\mathbf{\Phi}_{\text{noise}} (b) \in \mathbb{C} ^{M \times M}$, as follows:
\begin{equation}
\mathbf{\Phi}_{v} (b) = \sum\limits_{t=1}^T w_v(t,b)\mathbf{y}(t, b)\mathbf{y}(t, b)^\textrm{H} \textrm{ where } v \in \{\text{speech}, \text{noise}\},
\end{equation}
where $\mathbf{y}(t, b) \in \mathbb{C} ^{M}$ is an $M$-dimensional complex spectrum at time (frame) $t$ in frequency bin $b$.
$\mathbf{y}^\textrm{H}$ denotes the conjugate transpose.
$w_v(t,b) \in [0, 1]$ is the mask value.

The goal of GEV beamformer \cite{warsitz2007blind} is to estimate the beamforming filter $\mathbf{f} (b)$, which maximizes the expected SNR for each frequency bin $b$ as given by the equation below:
\begin{equation}
\mathbf{f}_{\textrm{GEV}} (b) = \argmax_{\mathbf{f}  (b)} \frac{\mathbf{f}^\textrm{H} (b)\mathbf{\Phi}_{\text{speech}} (b) \mathbf{f} (b)}{\mathbf{f}^\textrm{H}  (b) \mathbf{\Phi}_{\text{noise}} (b) \mathbf{f} (b)}.
\label{eq:maxmise_filter}
\end{equation}
Eq.~\eqref{eq:maxmise_filter} is equivalent to solve the following eigenvalue problem:
\begin{equation}
(\mathbf{\Phi}_{\text{noise}} (b) ) ^{-1} \mathbf{\Phi}_{\text{speech}} (b) \mathbf{f}  (b) = \lambda \mathbf{f}  (b),
\end{equation}
where $\mathbf{f}  (b) \in \mathbb{C} ^{M} $ at each frequency bin $b$ is the $M$-dimensional complex eigenvector and $\lambda$ is the eigenvalue. 

\begin{table*}[tbh]
  \caption{Speech Enhancement Scores}
  \label{tab:se_scores}
  \centering
  \scalebox{1.}{
  \begin{tabular}{ c c | c c c c | c c c c }
    \toprule
    \multicolumn{2}{c}{} & \multicolumn{4}{c}{\textbf{Dev (Simu)}} & \multicolumn{4}{c}{\textbf{Test (Simu)}} \\
   	Track & Enhancement Method & PESQ & STOI & eSTOI & SDR & PESQ & STOI & eSTOI & SDR \\
   	\midrule
    1ch & No Enhancement & 2.01 & 0.82 & 0.61 & 3.92 & 1.98 & 0.81 & 0.60 & 4.95 \\
    1ch & BLSTM Mask & \textbf{2.52} & \textbf{0.88} & \textbf{0.73} & \textbf{9.26} & \textbf{2.46} & \textbf{0.87} & \textbf{0.71} & \textbf{10.76} \\
       	\midrule
    2ch & BeamformIt & \textbf{2.15} & 0.85 & 0.65 & \textbf{4.61} & 2.07 & 0.83 & 0.62 & \textbf{5.60}  \\
    2ch & BLSTM Gev & 2.13 & \textbf{0.87} & \textbf{0.69} & 2.86 & \textbf{2.12} & \textbf{0.87} & \textbf{0.69} & 3.10\\
       	\midrule
    6ch & BeamformIt & 2.31 & 0.88 & 0.70 & \textbf{5.52} & 2.20 & 0.86 & 0.65 & \textbf{6.30} \\
    6ch & BLSTM Gev & \textbf{2.45} & 0.88 & \textbf{0.75} & 3.57 & \textbf{2.46} & \textbf{0.87} & \textbf{0.73} & 2.92 \\
    \bottomrule
  \end{tabular}
  }  
\end{table*}

\subsection{Time delayed neural network with lattice-free MMI}
\label{sec:tdnn}
For acoustic model, we use TDNN with LF-MMI training \cite{povey2016purely} instead of DNN+sMBR \cite{vesely2013sequence}. 
The architecture is similar to those described in \cite{peddinti2015time}. 
The LF-MMI objective function is shown below, which is different from usual MMI training \cite{povey2005discriminative} in a sense that we use phoneme sequence $L$ instead of a word sequence to narrow down a search space in the denominator:
\begin{equation}
 \mathcal{L}_{\text{MMI}} = \sum ^{N}_{n=1} \log \frac{p(\mathbf{O} ^{n} | S ^{n})^{\mathcal{C}} P(L ^{n})}{\sum _{L} p(\mathbf{O} ^{n} | S ^{L})^{\mathcal{C}} P(L) }
\end{equation}
where $p(\mathbf{O} ^{n} | S ^{L})$ is the likelihood function of a speech feature sequence $\mathbf{O} ^{n}$ given the state sequence $S ^L$ at $n$\textquotesingle th utterance. 
$P(L)$ is the phoneme language model probability and $\mathcal{C}$ is the probability scale.

Note that when combined with the data augmentation technique (described in Section \ref{sec:6data}), TDNN is more effective than DNN.

\subsection{LSTM language modeling}
\label{sec:lstm}
The LSTM based language model (LSTMLM) has been shown to be effective on language modeling \cite{sundermeyer2012lstm}. It is better in finding a longer period of contextual information than conventional RNN. With this property, LSTMLM can predict the next word in a more accurate way than RNNLM. Hence, instead of using a vanilla RNNLM \cite{mikolov2010recurrent}, we train an LSTMLM on WSJ data, which combines the use of subword features and one-hot encoding. An importance sampling method is used to speed up training.
Most important of all, a new objective function $\mathcal{L}_{\text{LM}}$ is used for LM training, 
which behaves like cross-entropy objective but trains
the output to auto-normalize in order to speed up test time
computation:
\begin{equation}
\mathcal{L}_{\text{LM}} = z_{j} + 1 - \sum_{i} \exp (z_{i})
\end{equation}
where $z$ is a pre-activation vector in the layer of neural network before the final softmax operation and $j$ is an index for the correct word.
More detail can be found in \cite{xuneural}.

\begin{table}[H]
  \caption{Experimental configurations}
  \label{tab:para}
  \centering
  \scalebox{1.}{
  \begin{tabular}{ l c }
    \toprule
    BLSTM mask estimation & \\
    \midrule
    input layer dimension & 513 \\
    L1 - BLSTM layer dimension & 256 \\
    L2 - FF layer 1 (ReLU) dimension & 513\\ 
    L3 - FF layer 2 (clipped ReLU) dimension & 513\\ 
    L4 - FF layer (Sigmoid) dimension & 1026\\
    $p_{dropout}$ for L1, L2 and L3 & 0.5\\
    \toprule
    TDNN acoustic model&  \\
   	\midrule
    input layer dimension & 40 \\
    hidden layer dimension & 750 \\
    output layer dimension & 2800 \\
    l2-regularize & 0.00005 \\
    num-epochs & 6 \\
    initial-effective-lrate & 0.003 \\
    final-effective-lrate & 0.0003 \\
    shrink-value & 1.0 \\
    num-chunk-per-minibatch & 128,64 \\
    \toprule
    LSTM language model &  \\
    \midrule
    layers dimension & 2048 \\
    recurrent-projection-dim & 512 \\
    N-best list size & 100 \\
	RNN re-score weight & 1.0 \\    
    \bottomrule
  \end{tabular}
  }
\end{table}

\section{Experiments}
\subsection{Speech Enhancement Experiments}
First experiments describe the speech enhancement performance of BLSTM-based speech enhancement.
For the single channel track, we used the BLSTM masking technique \cite{weninger2015speech} trained on the 6 channel data and took only the speech mask after the forward propagation.
We took a Hadamard product of the single channel spectrogram with the speech mask and used it as the enhanced signal to compare it with the original signal without any enhancement.
For the 2 channel and 6 channel tracks, we used the BLSTM based GEV beamformer described in Section \ref{sec:blstm} and compare it with BeamformIt. 
Four different scores as described in Section \ref{sec:intro} - PESQ, STOI, eSTOI and SDR are computed.
The BLSTM architecture used in the experiments is listed in Table \ref{tab:para}.

The enhancement scores are shown in Table \ref{tab:se_scores}. 
The 5th channel clean signal from the 6ch data convolved with room impulse response was used as the reference signal for computing all the four metrics. 
For the 1 channel track, the BLSTM mask gives significantly better scores in all four metrics compared to using the noisy data without any enhancement.
However, this is contrary to the ASR results, which will be discussed in the next section. 
BeamformIt has better SDR scores compared to BLSTM GEV in both the multi-channel tracks. 
Also, for both the multi-channel track data, eSTOI is slightly better for BLSTM GEV. 
In the 6ch track experiments, BLSTM GEV has a significantly better PESQ score.
Overall, BLSTM-based speech enhancement shows improvement in most of conditions except for the case of the multichannel SDR metric.

\subsection{Speech Recognition Experiments}
\label{sec:exp}
Our system is trained on the speech recognition toolkit Kaldi \cite{povey2011kaldi}. For TDNN acoustic model training, backstitch optimization method \cite{Wang2017} is used. The decoding is based on 3-gram language models with explicit pronunciation and silence probability modeling as described in \cite{chen2015pronunciation}. The model is re-scored by a 5-gram language model first. Then the Kaldi-RNNLM \cite{xuneural} is used for training the LSTMLM, and n-best re-scoring is used to improve performance. We got our best result in 6 channel experiments by averaging forward and backward LSTMLM. The RNN re-score weight is set to be 1.0, which means the results of 5-gram LM is completely discarded. All the results in this section are reported in terms of word error rate (WER). We also provide the parameters used in our system in Table~\ref{tab:para}.

\begin{table}[h]
  \caption{WER of adding enhanced data when using TDNN with BeamformIt and RNNLM in the 6 channel track experiment}
  \label{tab:enhan}
  \centering
  \begin{tabular}{ c c c c c }
    \toprule
    \textbf{Data Augmentation} & \multicolumn{2}{c}{\textbf{Dev (\%)}} & \multicolumn{2}{c}{\textbf{Test (\%)}} \\
	& real & simu & real & simu \\
   	\midrule
    all 6ch data & 3.97 & 4.33 & 7.04 & 7.39 \\
    all 6ch and enhanced data & \textbf{3.74} & \textbf{4.31} & \textbf{6.84} & 7.49 \\
    
    \bottomrule
  \end{tabular}
\end{table}

\begin{table*}[tbh]
  \caption{WER of 6 channel track experiments}
  \label{tab:6ch}
  \centering
  \begin{tabular}{ c c c c c c c c }
    \toprule
    \multicolumn{4}{c}{\textbf{Method}} & \multicolumn{2}{c}{\textbf{Dev (\%)}} & \multicolumn{2}{c}{\textbf{Test (\%)}} \\
   	Data Augmentation & Acoustic Model & Beamforming & Language Model & real & simu & real & simu \\
   	\midrule
    only 5th channel & DNN+sMBR & BeamformIt & RNNLM & 5.79 & 6.73 & 11.50 & 10.92 \\
    all 6ch data & DNN+sMBR & BeamformIt & RNNLM & 5.05 & 5.82 & 9.50 & 9.24 \\
    all 6ch and enhanced data & DNN+sMBR & BeamformIt & RNNLM & 5.62 & 6.46 & 10.27 & 9.41 \\
    all 6ch and enhanced data & TDNN with LF-MMI & BeamformIt & RNNLM & 3.74 & 4.31 & 6.84 & 7.49 \\
    all 6ch and enhanced data & TDNN with LF-MMI & BLSTM Gev & RNNLM & 2.83 & 2.94 & 4.01 & 3.80 \\
    all 6ch and enhanced data & TDNN with LF-MMI & BLSTM Gev & LSTMLM & \textbf{1.90} & \textbf{2.10} & \textbf{2.74} & \textbf{2.66} \\
    \bottomrule
  \end{tabular}
  
\bigskip
  \caption{WER of 2 channel track experiments}
  \label{tab:2ch}
  \centering
  \begin{tabular}{ c c c c c c c c }
    \toprule
    \multicolumn{4}{c}{\textbf{Method}} & \multicolumn{2}{c}{\textbf{Dev (\%)}} & \multicolumn{2}{c}{\textbf{Test (\%)}} \\
   	Data Augmentation & Acoustic Model & Beamforming & Language Model & real & simu & real & simu \\
   	\midrule
    only 5th channel & DNN+sMBR & BeamformIt & RNNLM & 8.23 & 9.50 & 16.58 & 15.33 \\
    all 6ch data & DNN+sMBR & BeamformIt & RNNLM & 6.87 & 8.06 & 13.33 & 12.57 \\
    all 6ch data & TDNN with LF-MMI & BeamformIt & RNNLM & 5.57 & 6.08 & 10.53 & 9.90 \\
    all 6ch and enhanced data & TDNN with LF-MMI & BeamformIt & RNNLM & 5.03 & 6.02 & 10.20 & 10.35 \\
    all 6ch and enhanced data & TDNN with LF-MMI & BLSTM Gev & RNNLM & 3.79 & 5.03 & 6.93 & 6.07 \\
    all 6ch and enhanced data & TDNN with LF-MMI & BLSTM Gev & LSTMLM & \textbf{2.85} & \textbf{3.94} & \textbf{5.40} & \textbf{5.03} \\
    \bottomrule
  \end{tabular}
  
\bigskip
  \caption{WER of 1 channel track experiments}
  \label{tab:1ch}
  \centering
  \begin{tabular}{ c c c c c c c c }
    \toprule
    \multicolumn{4}{c}{} & \multicolumn{2}{c}{\textbf{Dev (\%)}} & \multicolumn{2}{c}{\textbf{Test (\%)}} \\
   	Data Augmentation & Acoustic Model & Beamforming & Language Model & real & simu & real & simu \\
   	\midrule
    only 5th channel & DNN+sMBR & - & RNNLM & 11.57 & 12.98 & 23.70 & 20.84 \\
    all 6ch data & DNN+sMBR & - & RNNLM & 8.97 & 11.02 & 18.10 & 17.31 \\
    all 6ch data & TDNN with LF-MMI & - & RNNLM & 6.64 & 7.78 & 12.92 & 13.54 \\
    all 6ch data & TDNN with LF-MMI & - & LSTMLM & \textbf{5.58} & \textbf{6.81} & \textbf{11.42} & \textbf{12.15} \\
    all 6ch data & TDNN with LF-MMI & BLSTM masking & RNNLM & 13.15 & 15.62 & 22.47 & 21.61 \\
    all 6ch and enhanced data & TDNN with LF-MMI & BLSTM masking & LSTMLM & 6.78 & 9.10 & 13.64 & 14.95 \\
    \bottomrule
  \end{tabular}
\end{table*}

Table~\ref{tab:enhan} shows the effectiveness of the data augmentation for the system using TDNN with BeamformIt and RNNLM, which are described in Section \ref{sec:advance}, in the 6 channel track experiment. 
We confirmed the improvement by adding enhanced data in almost all cases except for the simulation test data. 
This is also found in 2 channels experiment when using TDNN (i.e. row 3 and row 4 in table~\ref{tab:2ch}).

Tables~\ref{tab:6ch} and \ref{tab:2ch} show the WER of 6 channel and 2 channel experiments. 
We change our experimental condition incrementally to compare the effectiveness of each method described in Section \ref{sec:advance}. 
In most of the situations, every method improved the WER steadily. 
We observed that the performance was degraded if we applied enhanced data on the system using DNN+sMBR (i.e. row 2 and row 3 in table~\ref{tab:6ch}), while TDNN with LF-MMI could make use of the enhanced data, as discussed above.
In addition, comparing with the speech enhancement results in Table \ref{tab:se_scores}, it shows that better speech enhancement scores do not necessarily gives lower WER. 
Especially, there always seems to be a negative correlation between the ASR performance and the SDR scores.

Table~\ref{tab:1ch} illustrates the results of the 1 channel track experiment. 
We found that BLSTM masking was not effective if we only used one microphone although it scores better in terms of all four speech enhancement metrics in Table \ref{tab:se_scores}.
From row 3 and row 5 of \ref{tab:1ch}, the WER with BLSTM masking was degraded more than twice when compared to the system without BLSTM masking. 
However, we also discovered that after adding the enhanced data into the system with BLSTM masking, the WER became closer to the best setup without masking, which can be seen in row 4 and row 6 of \ref{tab:1ch}.
Thus, adding the enhanced data seems to be a good strategy to mitigate the degradation of speech enhancement.

Finally, Table~\ref{tab:comparision} presents the comparison with the official baseline and top systems in the CHiME-4 challenge. 
We can see that all of these systems use a fusion technique to get their best WER. 
On the other hand, our proposed single system achieved 76\% relative improvement from the official baseline, and achieved the 2nd best performance.

\begin{table}[h]
  \caption{Final WER comparison for the real test set.}
  \label{tab:comparision}
  \centering
  \begin{tabular}{ c  c c }
    \toprule
    System & \# systems & WER (\%) \\
   	\midrule
    CHiME-4 baseline \cite{vincent2017analysis} & 1 & 11.51 \\
    Proposed system & 1 & 2.74 \\
    \midrule
    USTC-iFlytek\cite{du2016ustc} & 5 & 2.24 \\
    RWTH/UPB/FORTH\cite{menne2016rwth} & 5 & 2.91 \\
    MERL \cite{erdogan2016multi} & 6 & 2.98 \\
    \bottomrule
  \end{tabular}
\end{table}

\section{Conclusion}
This paper describes our single ASR system for CHiME-4 speech separation and recognition challenge. 
The system consists of BLSTM masked GEV beamformer (Section\ref{sec:blstm}), TDNN with LF-MMI as acoustic model (Section\ref{sec:tdnn}) and re-scoring using LSTMLM (Section\ref{sec:lstm}), which trained on all 6 channels data plus enhanced data generated by beamformer (Section\ref{sec:6data}). 
The system finally achieved 2.74\% WER, which outperforms the 2nd place result in the challenge.
The system is publicly available through the Kaldi speech recognition toolkit.

Our future work will explore different architectures for TDNN and LSTM networks to further improvement. 
Furthermore, this system can be applied to other multichannel tasks such as AMI \cite{mccowan2005ami}, and the CHiME-5 challenge \cite{chime2018asr}.

\bibliographystyle{IEEEtran}
\bibliography{mybib}


\end{document}